\documentstyle[multicol,aps,prb,psfig,epsfig]{revtex}
\def\zita{\zeta}
\def\Jnl#1#2#3#4{{#1} {\bf #2}, #3 (#4)}
\begin{document}
\title{Pulling hairpinned polynucleotide chains: Does base-pair stacking
interaction matter?}

\author{Haijun Zhou\cite{zhouhj}}

\address{
Max-Planck-Institute of Colloids and Interfaces,  
D-14424 Potsdam, Germany}

\author{Yang Zhang}

\address{ 
Institute of Theoretical Physics, The Chinese Academy of Sciences, 
P.O.Box 2735, Beijing 100080, China}

\date{\today}

\maketitle
\widetext

\begin{abstract}
Force-induced structural transitions both in relatively
random and in designed single-stranded DNA (ssDNA) chains are 
studied theoretically. At high salt conditions, ssDNA forms compacted
hairpin patterns stabilized by base-pairing and base-pair stacking
interactions, and a threshold external force is needed to pull  the
hairpinned structure into a random coiled one. 
The base-pair stacking interaction in the ssDNA chain makes this 
hairpin-coil conversion a discontinuous (first-order) phase 
transition  process characterized
by a force plateau in the force-extension curve,
while lowering this potential below 
some critical level turns this transition into  continuous (second-order)
type, no matter how strong the base-pairing interaction is. 
The phase diagram (including hairpin-I, -II, and random coil) 
is discussed as a function of 
stacking potential and external force.  
These results are in quantitative agreement with recent experimental 
observations of different ssDNA sequences, and they reveal the necessity
to consider the base-pair stacking interactions in order to understand
the structural formation of RNA, a polymer designed by nature itself.
The theoretical method used may be extended to study the
long-range interaction along double-stranded DNA caused  by  
the topological constraint of fixed linking number.  

\end{abstract}

\widetext
\begin{multicols}{2}
\section{Introduction}
\label{sect1}

Single-macromolecular manipulation techniques, such as atomic 
force microscopy and optical tweezer methods, were recently used by
many authors to investigate the mechanical properties of double-stranded
DNA (dsDNA) and to understand how the tension in 
dsDNA influences the interactions between
DNA and proteins such as DNA/RNA polymerases, type II topoisomerases,
and RecA proteins (for a brief review see Ref.~[1]).
Many useful insights have been obtained through a detailed analysis of 
the experimental data generated by these precise  measurements  (see, 
for example, Refs.~[2-9]).
These new experimental techniques were recently
used also  on RNA and 
single-stranded DNA (ssDNA) to explore the structure
formation of  these polynucleotides.\cite{bustamante2000,smith1996,wuite2000,maier2000}
A RNA or ssDNA is a linear
chain of nucleotide bases. At physiological conditions 
(concentration $\sim$ $0.1$M Na$^+$ and temperature $\sim$ $300$K),
RNA molecules in biological cells can fold into  stable native
configurations (as in the case of proteins) to fulfill  their various
biological  functions.\cite{watson1987}   
The higher-order structures in RNA and ssDNA are
caused by the possibilities to form base-pairs between the complementary
nucleotide bases,  A and U (T, in the case of ssDNA), and  G and C. These
structures can be further stabilized by the vertical
stacking interactions between
nucleotide base-pairs. In a long polynucleotide chain, because there are a
great many different  ways to form base-pairs, the accessible configurational
space is very large, and the equilibrium configurations are
determined
by the competition between entropy and the interaction energy.
To investigate how base-pairing and base-pair
stacking  influence  structure formations of polynucleotide is
therefore of both biological and theoretical interest.

Recent measurements demonstrated that the structures of polynucleotides 
are sensitive to the ionic strength of the aqueous  
solution.\cite{bustamante2000,smith1996,wuite2000,maier2000}
In a high salt environment (for example, 150 mM Na$^+$ or 
5 mM Mg$^{2+}$), electrostatic repulsive forces 
between the negatively-charged nucleotide phosphate groups 
are largely shielded and base-pairing is favored; 
furthermore, experimental observations
suggested the formation of hairpinned configurations in ssDNA.\cite{bustamante2000} 
To understand the hairpin formation in ssDNA and RNA, 
Montanari and M\'{e}zard suggested very recently a model for
these  polynucleotides.\cite{montanari2000}
Base-pairing interaction is considered in their model, 
and their theoretical calculation can  reproduce the experimental
force-extension curve measured on a ssDNA chain whose  nucleotide bases 
are arranged in relatively random order. 
It can be shown (see below and also Ref.~[14]) that the force-induced structural 
transition in  ssDNA is a second-order continuous phase transition process, 
characterized by a gradual decrease in  the number of base-pairs as the external force
is increased beyond a certain  threshold value. 

On the other hand, Rief {\it et al.}\cite{rief1999} observed  a  quite different 
phenomenon: when they pulled a ssDNA chain made of poly(dA-dT) or poly(dG-dC)
sequence apart with an atomic force microscope, 
they found that  the distance between its two ends suddenly changes
from nearly zero to a  value comparable to its total contour length during
a very  narrow force range.

A question thus arises: Why is the force-induced structural transition 
of ssDNA gradual and continuous in one case and  highly cooperative in the
other case, even though the solution conditions in these two kinds of
experiments are comparable? 
We noticed that there may be strong base-pair stacking interactions
in the designed ssDNA chains, while stacking interactions are   
absent or negligible in the relatively random sequences.
We have previously shown that the short-ranged base-pair
stacking potential in dsDNA dominates the energy 
contribution\cite{hagerman1988} and it 
is the reason both for the stability of dsDNA and
its high extensibility at large forces.\cite{zhou1999,zhou2000} 
Is the dramatically different  experimental observation of 
Rief {\it et al.}\cite{rief1999} (compared with those of 
[1,11,12])  caused by base-pair stacking interaction? Given the fact that the  
stacking potential  is a kind of short-ranged (nearest-neighbor) interaction,
does the cooperative transition reported in [15] 
for such an 1D system really differ qualitatively from  that
observed  in [1,11]?  

To address these questions, we extend the work of Montanari 
and M\'{e}zard\cite{montanari2000} to include the effect of
stacking interactions, and thus obtain a more realistic  model 
for polynucleotides. We find that when the base-pair stacking
potential is small, the ssDNA chain is in a hairpinned macro-state with the 
base-pairs only weakly stacked (hairpin-I), while when the stacking potential
is large, the ssDNA is in a hairpinned macro-state with almost 
complete pairing and stacking (hairpin-II). The force-induced transition 
between hairpin-II and random coil is first-order, while the
transition  between hairpin-I and random coil is continuous.  
There is also a continuous phase  transition between 
hairpin-I and hairpin-II, which may be induced by  changing 
the nucleotide arrangement or by environmental
variations in temperature or ionic strength. 
A simplified treatment of the same question was reported 
earlier.\cite{zhou2001}

The paper is organized as following:
Section \ref{sect2} introduces the theoretical  model of ssDNA;
and Sec.~\ref{sect3} presents some qualitative
arguments concerned with the structural transitions in the model of ssDNA;
in Sec.~\ref{sect4} we perform  a direct comparison with 
experimental observations.
The main points of the paper are summarized in Sec.~\ref{sect5}, accompanied
by some qualitative discussions on  RNA secondary structure formation.
We suggest the possibility to extend the theoretical method in this paper
to address the long-range interactions along topologically constrained
double-stranded DNA molecules, i.e., dsDNA with fixed linking number.   

\section{Model of ssDNA with base-pair stacking interaction}
\label{sect2}

The model used in the present study is similar with  that developed earlier  
by Montanari and M\'{e}zard,\cite{montanari2000} 
albeit with base-pair stacking potential included.  We discuss a fictitious
polymer chain made of $N$ tiny beads (bases) with index $i$ from $1$ to $N$
(each base in the model may correspond to several
bases in a realistic polynucleotide).
Between any two adjacent bases ($i$ and $i+1$) there is an elastic bond 
of equilibrium length $b$ and length variance
$\ell$ ($\ell \ll b$, i.e., the bond is considerably rigid). For any
two bases $i$ and $j$, if their mutual distance $|{\bf r}_{i j}|$
is less than an interaction radius $a$ there could
be a pairing potential $V_{\rm pair}^{i,j} ({\bf r}_{i j})$ (and we say
that an $i$-$j$ base-pair is formed);
and if any two base-pairs are nearest neighbors to
each other (that is, $i$-$j$ and $(i+1)$-$(j-1)$),
there is an additional  stacking interaction
$J_{i,i+1}^{j,j-1} ({\bf r}_{i j}, {\bf r}_{i+1,j-1})$. 
In this work, as what is assumed in other previous theoretical 
treatments,\cite{higgs1996,bundschuh1999,pagnani2000} 
only the secondary pairing patterns are
considered, which leads to the following constraints on 
the pairing between bases: (1) each base can be unpaired or be 
paired to at most one another base;  (2) if base $i$ is paired to base
$j$ (suppose $i<j$) and $k$ to $l$ ($k<l$), then either $i<j<k<l$
($``$independent") or
$i<k<l<j$ ($``$nested").  
 
Because of these restrictions, the partition function can be calculated  
iteratively.\cite{montanari2000,higgs1996,bundschuh1999} 
In the case when the pairing interaction radius $a$ is 
far less than  bond equilibrium length $b$ ($a\ll b$), 
the total statistical weight $Z_{j,i}({\bf r})$
for a polynucleotide segment (from base $i$ to $j$) whose ends being
separated by a distance ${\bf r}$
is expressed as  
\end{multicols}
\widetext
\begin{eqnarray}
 & &Z_{j,i}({\bf r})=\int d{\bf u} \mu ({\bf u}) Z_{j-1, i}({\bf r}-{\bf u})
+f_{j i}({\bf r}) \int d{\bf u}_1 d{\bf u}_2 \mu({\bf u}_1) \mu({\bf u}_2)
Z_{j-1,i+1}({\bf r}-{\bf u}_1-{\bf u}_2) \nonumber \\
 & &\;\;\;\; +\sum\limits_{k=i+1}^{j-2} \int d{\bf u}_1  d{\bf u}_2 
d{\bf u}_3 d {\bf v} \mu({\bf u}_1) \mu({\bf u}_2) \mu({\bf u}_3)
f_{j k}({\bf v}) Z_{k-1,i}({\bf r}-{\bf u}_1 -{\bf v}) Z_{j-1,k+1}({\bf v}-{\bf
u}_2-{\bf u}_3) \nonumber \\
 & &\;\;\;\;+s(a-|{\bf r}|)\exp(-\beta V_{\rm pair}^{j,i} ({\bf r}))
\int d{\rm u}_1  d{\rm u}_2 \mu({\bf u}_1) \mu({\bf u}_2)  
g_{i,i+1}^{j,j-1}({\bf r}, {\bf r}-{\bf u}_1-{\bf u}_2)
Z_{j-1,i+1}^{(p)} ({\bf r}-{\bf u}_1-{\bf u}_2) \nonumber \\
 & &\;\;\;\;+\sum\limits_{k=i+1}^{j-2} \int d {\bf u}_1 
d{\bf u}_2 d {\bf u}_3  d {\bf v} s(a-|{\bf v}|)
\exp(-\beta V_{\rm  pair}^{j,k}({\bf v}))
g_{k,k+1}^{j,j-1}({\bf v},{\bf v}-{\bf u}_2-{\bf u}_3)\times \nonumber \\ 
 & &\;\;\;\;\;\;\;\;\;\;\;\;\;\;\;\;\;
Z_{k-1,i}({\bf r}-{\bf u}_1 -{\bf v}) Z_{j-1,k+1}^{(p)}
({\bf v}-{\bf u}_2-{\bf u}_3). 
\label{eq:eq01}
\end{eqnarray}
\widetext
\begin{multicols}{2}
Here,
$ 
\mu({\bf r}) \propto \exp(-{(|{\bf r}|-b)^2/2 \ell^2})
$
is the probability density for the bond vector  ${\bf r}$; 
$
f_{j i} ({\bf r})=\exp[-\beta V_{\rm pair}^{j,i} ({\bf r}_{i j})] -1
$
characterizes  the pairing interaction between bases $i$ 
and $j$ (this viral expansion form $f$ is introduced because that,
in the first integral on the right-hand side of the equality in 
Eq.~(\ref{eq:eq01}), an vanishing pairing potential between 
the end base $j$ and any another base is assumed and
the contributions  of all these  possible base-paired configurations are included.
Such spurious contributions should be removed since the pairing potential is 
actually nonzero); 
$s(x)$ is a signal function, $s(x)=1$ if $x\ge 0$ and $0$ otherwise;
$
g_{i, i+1}^{j,j-1}({\bf r}_{i,j},{\bf r}_{i+1,j-1})=\exp[-\beta J_{i, i+1}^{j, j-1}
({\bf r}_{i j}, {\bf r}_{i+1, j-1})]-1
$
characterizes  the base-pair stacking interaction (the reason for the
introduction of the viral coefficient $g$ can be similarly understood as  
that for the introduction of the  viral coefficient $f$). 
$Z_{j,i}^{(p)}$ is the statistical weight for a ssDNA segment 
whose two end bases ($i$ and $j$) forming a base pair: 
\end{multicols}
\widetext
\begin{eqnarray}
Z_{j,i}^{(p)} ({\bf r}) &=&s(a-|{\bf r}|) \exp(-\beta V_{\rm pair}^{j,i}
({\bf r})) \int d {\bf u}_1 d {\bf u}_2 \mu ({\bf u}_1) \mu ({\bf u}_2)
Z_{j-1,i+1} ({\bf r}-{\bf u}_1 -{\bf u}_2) \nonumber \\
 &+&s(a-|{\bf r}|) \exp(-\beta V_{\rm pair}^{j,i}({\bf r}))
\int d {\bf u}_1 d {\bf u}_2 \mu({\bf u}_1)\mu({\bf u}_2) 
g_{i,i+1}^{j,j-1}({\bf r}, {\bf r}-{\bf u}_1 -{\bf u}_2)\times \nonumber \\
& &\;\;\;\;\;\;\;\;\;\;
 Z_{j-1,i+1}^{(p)}({\bf r}-{\bf u}_1 -{\bf u}_2).
\label{eq:eq02}
\end{eqnarray}

\widetext
\begin{multicols}{2}
The statistical property of this model system is determined by
the coupled equations (\ref{eq:eq01}) and (\ref{eq:eq02}).
In the general case where  the pairing and stacking potentials
are dependent on base index, it is certainly of no hope to 
investigate the property of the system analytically. Here
we simplify our task by  assuming that  the pairing and 
stacking potentials between any  bases have the same form, i.e.,
assuming the polymer chain to be homogeneous.\cite{note:no1} 
 Then in Eqs.~(\ref{eq:eq01}) and 
(\ref{eq:eq02}) all the subscripts specifying 
specific bases can be dropped, especially we can write $f_{ji}({\bf r})$ as
$f({\bf r})$,  $Z_{j,i}({\bf r})$ as $Z_{j-i}({\bf r})$ and
$Z_{j,i}^{(p)}({\bf r})$ as $Z_{j-i}^{(p)}({\bf r})$.
In the model, the stacking potential is a function of the inter-base
distances for  the two base-pairs concerned.  
Since the base-pair's inter-base distance  is usually very small 
(less than $a$), 
we can approximate the stacking potential to be constant
in this range, and hence we just denote the
stacking potential to be a constant $J$ and denote 
$g=g_{i,i+1}^{j,j-1}({\bf r}_{i,j},{\bf r}_{i+1,
j-1})=const.$ 

The Fourier transforms of the generating functions (Laplace transforms) 
of the statistical weights are defined as
\begin{equation}
\Xi(\zita, {\bf p})= \int d{\bf r}\left(\sum\limits_{n=0}^{\infty}
Z_{n}({\bf r}) \zita^n\right) \exp(i {\bf p}\cdot {\bf r}),
\label{eq:eq03}
\end{equation}
and
\begin{equation}
\Xi^{(p)} (\zita,{\bf p}) = \int d{\bf r}
\left(\sum\limits_{n=0}^{\infty}Z_{n}^{(p)} ({\bf r}) \zita^n\right)
\exp(i {\bf p}\cdot {\bf r}).
\label{eq:eq04}
\end{equation}
Considering the iterative expressions for $Z_{n}$ and $Z_{n}^{(p)}$ in
Eqs.~(\ref{eq:eq01}) and (\ref{eq:eq02}), we can show that
\end{multicols}
\widetext
\begin{eqnarray}
\Xi(\zita,{\bf p})&=&[1+\zita \sigma({\bf p})\Xi(\zita,{\bf p})]\times \nonumber \\
 & &\left(1+{\zita^2 b^3 \over (2 \pi)^3} \left[\int d {\bf q}\sigma^2({\bf q}) 
\gamma({\bf p}-{\bf q})\Xi(\zita,{\bf q}) 
+ g \int d {\bf q} \sigma^2 ({\bf q}) \gamma^{\prime}({\bf p}-{\bf q})
\Xi^{(p)} (\zita, {\bf q})\right]\right),
\label{eq:eq05}
\end{eqnarray}
and 
\begin{equation}
\Xi^{(p)} (\zita,{\bf p}) = {\zita^2 b^3 \over (2\pi)^3}
\int d{\bf q} \sigma^2 ({\bf q}) \gamma^{\prime} ({\bf p}-{\bf q})
\left[\Xi(\zita, {\bf p})+g \Xi^{(p)} (\zita, {\bf q})\right],
\label{eq:eq06}
\end{equation}
\widetext
\begin{multicols}{2}
where $\sigma({\bf p})=\int d {\bf r} \mu({\bf r}) \exp(i {\bf p}\cdot 
{\bf r})=(\sin{p b}/pb)\exp(-p^2 l^2/2)$, with $\sigma({\bf 0})=1$;  
$\gamma({\bf p})=b^{-3}\int_{|{\bf r}|\le a} 
d {\bf r} f({\bf r}) \exp(i {\bf p}\cdot {\bf r})$
and $\gamma^{\prime}({\bf p})=\gamma({\bf p})+(a/b)^3$. Since $a \ll b$, we
have $\gamma^{\prime}=\gamma$ and both $\gamma$ and $\gamma^{\prime}$
can be regarded as independent of momentum ${\bf p}$. 
Then Eqs.~(\ref{eq:eq05}) and (\ref{eq:eq06}) lead to
\begin{equation}
\Xi(\zita,{\bf p})={D(\zita) \over 1-\sigma({\bf p}) D(\zita)}, \label{eq:eq07}
\end{equation}
where
\begin{equation}
D(\zita)=\zita +\zita \Xi^{(p)} (\zita)=-\eta_1 \zita^3 +\eta_2 \zita^2
+\zita, \label{eq:eq08}
\end{equation}
with  coefficient $\eta_1$  a constant and $\eta_2$  related to $D$:
\begin{eqnarray}
& & \eta_1 = g \gamma b^3(2 \pi)^{-3} \int d {\bf q} \sigma^2 ({\bf q})
=g \gamma (4\pi)^{-3/2} (b/\ell), 
\label{eq:eq09} \\
& & \eta_2(D)=D \eta_1 \left[1 + (b^2 \ell/g \pi^{3/2}) 
\int d {\bf q} {\sigma^2 ({\bf q})\over 1-\sigma({\bf q}) D} \right].
\label{eq:eq10} 
\end{eqnarray}
Equations (\ref{eq:eq07}) and (\ref{eq:eq08}) 
are the central equations of this article. 

When a external force ${\bf F}$ is applied at the end of the ssDNA chain,
the total partition function is 
%\begin{equation}
$
Z_{N}^{F} =\int d{\bf r} Z_{N}({\bf r}) \exp(-\beta {\bf F} \cdot {\bf r})
%\label{eq:eq11}
$.
%\end{equation}
The Laplace transform of this partition function is calculated to be 
\begin{equation}
\sum\limits_{N=0}^{\infty} Z_N^F \zita^N =
\Xi(\zita, -i\beta {\bf F}),
\label{eq:eq12}
\end{equation}
where $\Xi(\zita,-i\beta {\bf F})$ is determined by Eq.~(\ref{eq:eq07}).

For a linear  polymer system, the free energy is  an extensive
quantity proportional to the number of monomers $N$ in the thermodynamic 
limit. This indicates that the smallest
positive singularity point of the function $\Xi(\zita, -i\beta {\bf F})$
in the variable $\zita$ corresponds to the free energy density of the 
ssDNA chain.\cite{lifson1964,poland1966}  This point is  used in 
the next section.

\section{Qualitative analysis of the hairpin-coil transition}
\label{sect3}

It is evident   that  $\Xi(\zeta, -i\beta {\bf F})$ has a 
pole $\zeta_{\rm pole}$ determined
by  
\begin{equation}
D(\zeta_{\rm pole})=1/\sigma(-i\beta {\bf F}) =
{\beta F b \over \sinh(\beta F b)} \exp( -\beta^2 F^2 \ell^2 /2),
\label{eq:eq13} 
\end{equation}
where $F=|{\bf F}|$. The function at the right side of the equality
monotonously decreases with $F$  from  $1$ at $F=0$ to $0$ as 
$F\rightarrow
\infty$.  On the other hand,  the function  $D(\zeta)$
 is related to the Laplace transform of
the hairpinned configurations as indicated by Eq.~(\ref{eq:eq08}),
therefore $D(\zeta)$  has  a finite convergence radius, and it is
an increasing function of $\zeta$ before this radius
is reached. The singularity  of $D(\zeta)$ is related to
the roots of Eq.~(\ref{eq:eq08}).   Although analytical expressions for the 
roots of this third-order equation are available,
they are lengthy and here we discuss the statistical property of the 
system through an alternative routine.
First we consider two   extreme cases:

{\em Case A: the base-pair stacking potential $J=0$}. 
\hspace{1.0cm} In this case $g=0$ and $\eta_1=0$ 
and Eq.~(\ref{eq:eq08}) reduces to second-order. 
This situation  has been studied by Montanari and M\'{e}zard\cite{montanari2000}
and they found that $D(\zeta)$ has a second-order branching point
at $\zeta_{\rm bp}$ equaling to the maximum  of the expression 
$(-1+\sqrt{1+4 D \eta_2})/2 \eta_2$, which is reached at 
$D=D_{\rm bp} <1$.
When the external force is less than the threshold value 
$F_{\rm cr}$ determined by Eq.~(\ref{eq:eq13}) at $D=D_{\rm bp}$, 
the polymer  resides in the hairpinned phase with zero 
extension, and the free energy density equals to $\phi(F)=(1/\beta) \ln
\zeta_{\rm bp}$ and is force-independent. 
At $F=F_{\rm cr}$ there is a second-order
continuous hairpin-coil phase-transition 
(because $d\zeta/d D =0$ at $D_{\rm bp}$), 
and the free energy
density is changed to  $\phi(F)=(1/\beta) \ln (\zeta_{\rm pole})$.

{\em Case B: the  stacking potential $J$ so large that $g\gg 1$.}
\hspace{1.0cm} In this case,  Eq.~(\ref{eq:eq10}) indicates that
$\eta_2=D \eta_1$ and Eq.~(\ref{eq:eq08}) is equivalent to 
\begin{equation}
(\zeta+1/\sqrt{\eta_1})(\zeta-D)(\zeta-1/\sqrt{\eta_1})=0.
\label{eq:eq14}
\end{equation}
We readily see that when $D\le 1/\sqrt{\eta_1}$ 
the smallest positive root of this equation is $\zeta
=D$; and when $D> 1/\sqrt{\eta_1}$, the smallest positive solution is a
constant $\zeta=1/\sqrt{\eta_1}\propto (g\gamma)^{-1/2}$.
Most importantly, we have $d\zeta/d D=1$ as $D$ approaches 
$1/\sqrt{\eta_1}$ from below. As a consequence,
for $F< F_{\rm cr}$ which is determined by Eq.~(\ref{eq:eq13})
with $D=1/\sqrt{\eta_1}$, the polymer  is 
in the hairpinned state and the free energy density is
$\phi(F)=-(1/2\beta)\ln \eta_{1}
\propto -(1/2\beta)\ln (g \gamma)$ (again independent of $F$). 
For $F> F_{\rm cr}$ the system is in the random
coil state, and $\phi(F)=(1/\beta) \ln (\zeta_{\rm pole})$. Here
$\zeta_{\rm pole}$ is determined by Eq.~(\ref{eq:eq13}) (with  $D=\zeta_{\rm pole}$). 
At $F=F_{\rm cr}$ there is a 
{\em first-order} hairpin-coil phase-transition, resulted from
the fact that $d\zeta/d D|_{D=1/\sqrt{\eta_1}}=1$.  

\begin{figure}
\center
\psfig{file=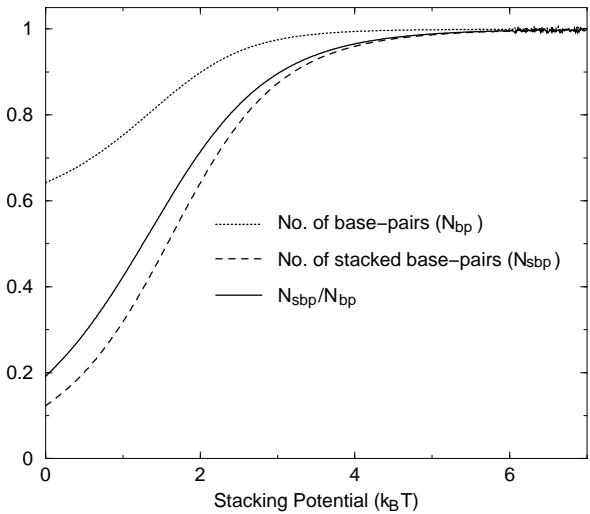,width=8.0cm}
\caption{The relationship between the order parameter
$N_{\rm sbp}/N_{\rm bp}$ and the effective
stacking potential.
Here $N_{\rm bp}$ is the total number of 
paired bases (in units of $N/2$), and
$N_{\rm sbp}$ is  the total number of stacked base-pairs
(also in units of $N/2$). 
In the numerical  calculation of this figure, 
$\gamma$ (it accounts for the pairing intensity) is set to $1.9$  
and external force $F=0$.
\label{fig:fig01}}
\end{figure}

Comparing case A and case B, we have the impression that 
the inclusion of base-pair
stacking interaction may dramatically change the statistical property
of the ssDNA system, even the order of the hairpin-coil phase transition.
This is understandable qualitatively. The  stacking interaction has two 
effects: (1) it makes base-pairing even more favorable; and (2) it
causes the formed base-pairs to aggregate into large stacked blocks.
Since the order of the  hairpin-coil transition is related to the
intensity of the stacking potential,  the hairpin macro-state at 
low stacking intensity is anticipated to be different to the
hairpin macro-state at high stacking intensity, and a continuous 
phase transition between these two hairpin macro-states can be 
predicted as the average base-pair stacking intensity changes. 
This insight is confirmed by observing how the order parameter of
this system changes with stacking potential, as shown in Fig.~\ref{fig:fig01}.
Later we will refer to the hairpin macro-state at low 
stacking intensity as hairpin-I and that at high stacking intensity
as hairpin-II.       

In the general case, the  average number of base-pairs 
(in units of $N/2$) is calculated according to
%\begin{equation}
$
N_{\rm bp}=-2 {\partial \phi/\partial \ln\gamma},
$
%\label{eq:eq15}
%\end{equation}
and the average number of stacked base-pairs (also in units of $N/2$) is
%\begin{equation}
$
N_{\rm sbp}=-2{\partial \phi/\partial\ln g},
$
%\label{eq:eq16}
%\end{equation}
and the relative extension of each ssDNA  bond along the direction of
the external force is obtained by
$
Ex=-{\partial \phi / \partial F}.
$

We demonstrate in Fig.~\ref{fig:fig01} how the values of $N_{\rm bp}$,
$N_{\rm sbp}$, and the ratio $N_{\rm sbp}/N_{\rm bp}$ change with 
strength of base-pair
stacking potential. In this figure $\gamma$  is fixed to $1.9$  (just serves
as an example) and the external force is set to zero. 
We see that as the stacking potential $J$ increases, all 
these three quantities increases and they approaches $1$ as $J$ reaches
about $J_{\rm cr} \simeq 6 k_B T$. The ratio $N_{\rm sbp}/N_{\rm bp}$ could
serve as an order parameter. Figure \ref{fig:fig01} shows that when
stacking potential is high, almost all the nucleotide bases 
are paired and  stacked; 
lowering stacking potential from $J>J_{\rm cr}$ to $J<J_{\rm cr}$
there is a continuous phase transition where 
such a highly stacked configuration is gradually melted out and a
growing fraction of bases becomes unpaired and unstacked.
This is a structural transition
between a highly stacked hairpin macro-state  (hairpin-II)
and a  loosely stacked (or irregular)  hairpin macro-state 
(hairpin-I). This hairpin I-II transition  is induced by changing 
the average effective base-pair stacking interaction in the polymer.    
Giving a polynucleotide chain, macroscopically
the effective base-pairing interaction is  determined mainly
by its base composition, while the effective   stacking potential
is determined mainly by  the particular arrangement of the bases
along the chain. Hence, for different polynucleotides with the same
base composition, the folded hairpin configurations may be quite
different and fall into two gross catalogue, dependent on their
particular base sequence arrangement. This is consistent with
the observation that certain RNA chains have unique stable 
native configurations. The transition between hairpins-I and -II can
also be induced by temperature changes or variations in
solution ionic concentrations.   

It should be mentioned that the critical stacking potential $J_{\rm cr}$ 
is not sensitive to the pairing potential. This can be seen from the
fact that the hairpin I-II transition occurs when 
the second term in the square brackets of Eq.~(\ref{eq:eq10}) approaches 
zero, a condition which is  solely satisfied when the stacking potential
(and hence $g$) becomes large.  

\begin{figure}
\center
\psfig{file=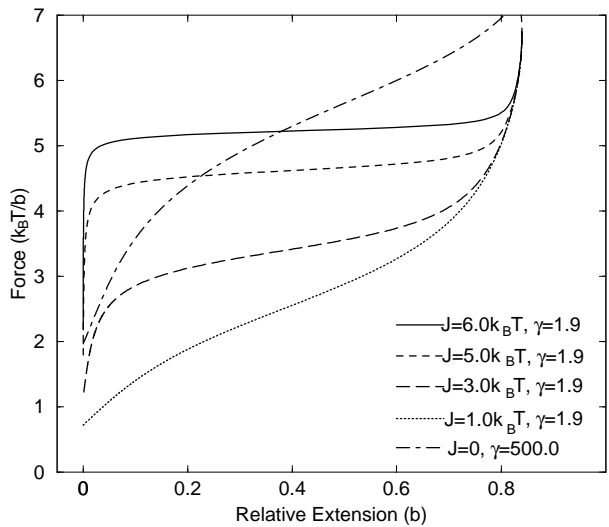,width=8.0cm}
\caption{The relationship between the average end-to-end extension of
a polynucleotide and the applied force at 
different effective base-pairing  and base-pair
stacking intensities. Here $b$ denotes the Kuhn length of
a ssDNA segment,  $\gamma$ accounts for the intensity of the
pairing potential between  nucleotide bases,  and 
$J$ is the effective stacking potential between nucleotide base-pairs. 
\label{fig:fig02}}
\end{figure}

The force-induced
hairpin-coil transition is then second-order or first-order depending 
on whether the hairpinned configuration is type I or type II. 
In Fig.~\ref{fig:fig02} the force vs extension relationship for
ssDNA at different pairing and 
stacking interaction intensities are shown.  As is 
expected, at large stacking intensity, there is a broad
force-plateau and this force-plateau disappears as stacking potential is 
lowered.  It is striking to notice that only the inclusion of base-pair
stacking interaction can lead to the appearance of a highly cooperative
hairpin-coil transition. As shown also in Fig.~\ref{fig:fig02} (the dot-dashed
curve), if $J=0$ the force-extension curve is always continuous and gradual 
no matter how strong the base-pairing interaction is.

\begin{figure}
\center
\psfig{file=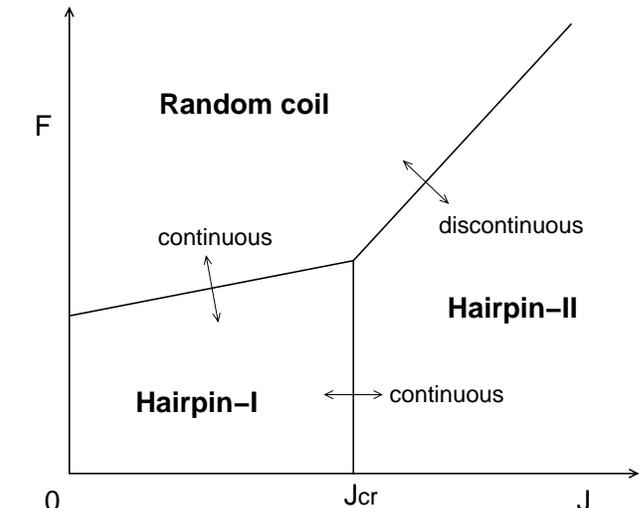,width=8.0cm}
\caption{The qualitative phase-diagram of a polynucleotide
at high-salt conditions. The phases hairpin-I and hairpin-II 
are defined in the main text. $F$ denotes the
intensity of the external force.   \label{fig:fig03} }
\end{figure}

These theoretical predictions are summarized in the qualitative phase-diagram
depicted in Fig.~\ref{fig:fig03}.

\section{Quantitative comparison with experimental observations}
\label{sect4}

In the last section we have analyzed the main qualitative predictions
of the present polynucleotide elastic model.  In this section we
apply these results to attain a quantitative understanding of  some
reported experimental findings.

\begin{figure}
\center
\psfig{file=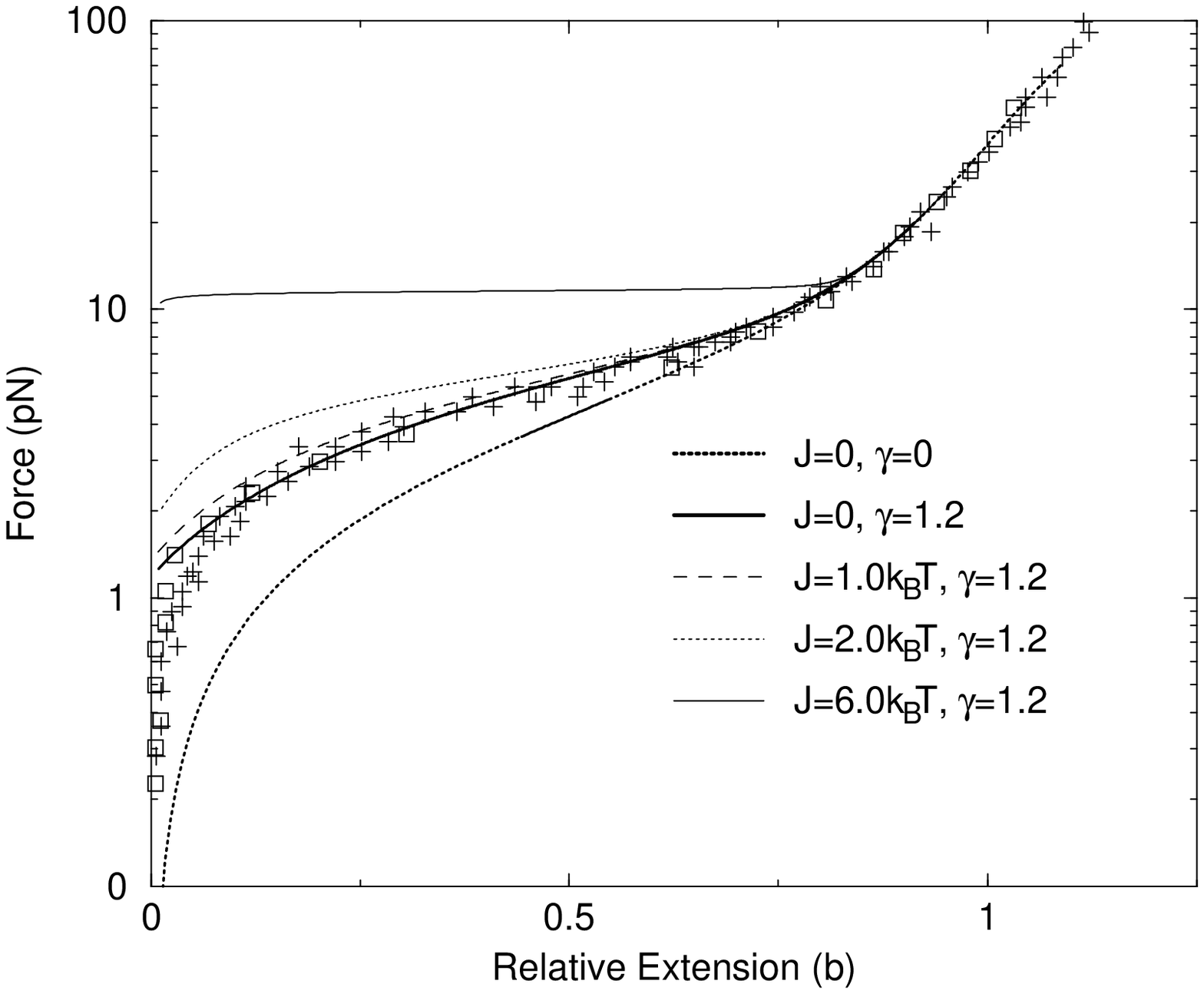,width=8.0cm}
\caption{Force vs extension curves for a  ssDNA polymer with
its bases randomly ordered. The experimental data come from 
Ref.~[1] (pluses) and
[12] (squares), and  the  ionic concentration is
$5$mM Mg$^{2+}$.  The theoretical
calculation are performed with $b=1.7$nm and $\ell=1.105$$\AA$, and the other
two fitting parameters ($\gamma$ and $J$) are 
listed in the figure.     \label{fig:fig04}}
\end{figure}

Bustamante {\it et al.} and others
\cite{bustamante2000,wuite2000,maier2000}
have pulled a plasmid ssDNA fragment of 10.4 kilobases  under 
different ionic concentrations. They found that the elastic response
of ssDNA deviates from that of a random coil both at high and at
low ionic conditions. It was suspected that in high salt environments
there is partial hairpin formation in the ssDNA 
chain.\cite{bustamante2000,wuite2000}
As is consistent with this insight,
Fig.~\ref{fig:fig04} demonstrates that at high salt conditions, the
ssDNA made of such a relatively random sequence can be well modeled by the
present model with an effective base-pairing interaction
characterized by the dimensionless quantity $\gamma=1.2$ and Kuhn length
$b=1.7$nm and bond length variation $\ell =0.065 b=1.105$$\AA$.
The effective base-pair stacking interaction for radom sequences 
seems to be quite small and is set to $J=0$ in the fitting (the thick solid line
in Fig.~\ref{fig:fig04}). Such a 
comparison has already been performed by Montanari and M\'{e}zard
earlier.\cite{montanari2000} 
For random sequences the reason that the effective stacking interaction
is negligible may be explained as follows: In the polynucleotide chain
whose bases arrange randomly, the formed base-pairs are usually
seperated from each other and the possibility for two base-pairs to
be adjacent to each other is quite small. 

Even when the effective stacking potential $J$ between two consective
node-pairs becomes comparable with thermal energy $k_B T$ (the thin doshed
line in Fig.~\ref{fig:fig04}), the gradual force-extension pattern
is only slightly changed. It indicates that the stacking interaction still plays a less
important role than the pairing interaction. But the fitting 
deteriarates much for $J\ge 2 k_B T$ if the pairing potential is
kept constant (thin dotted line): for
example, at $J=6 K_B T$ (thin solid line) there is a wide
plateau in the theoretical force-extension curve.

\begin{figure}
\center
\psfig{file=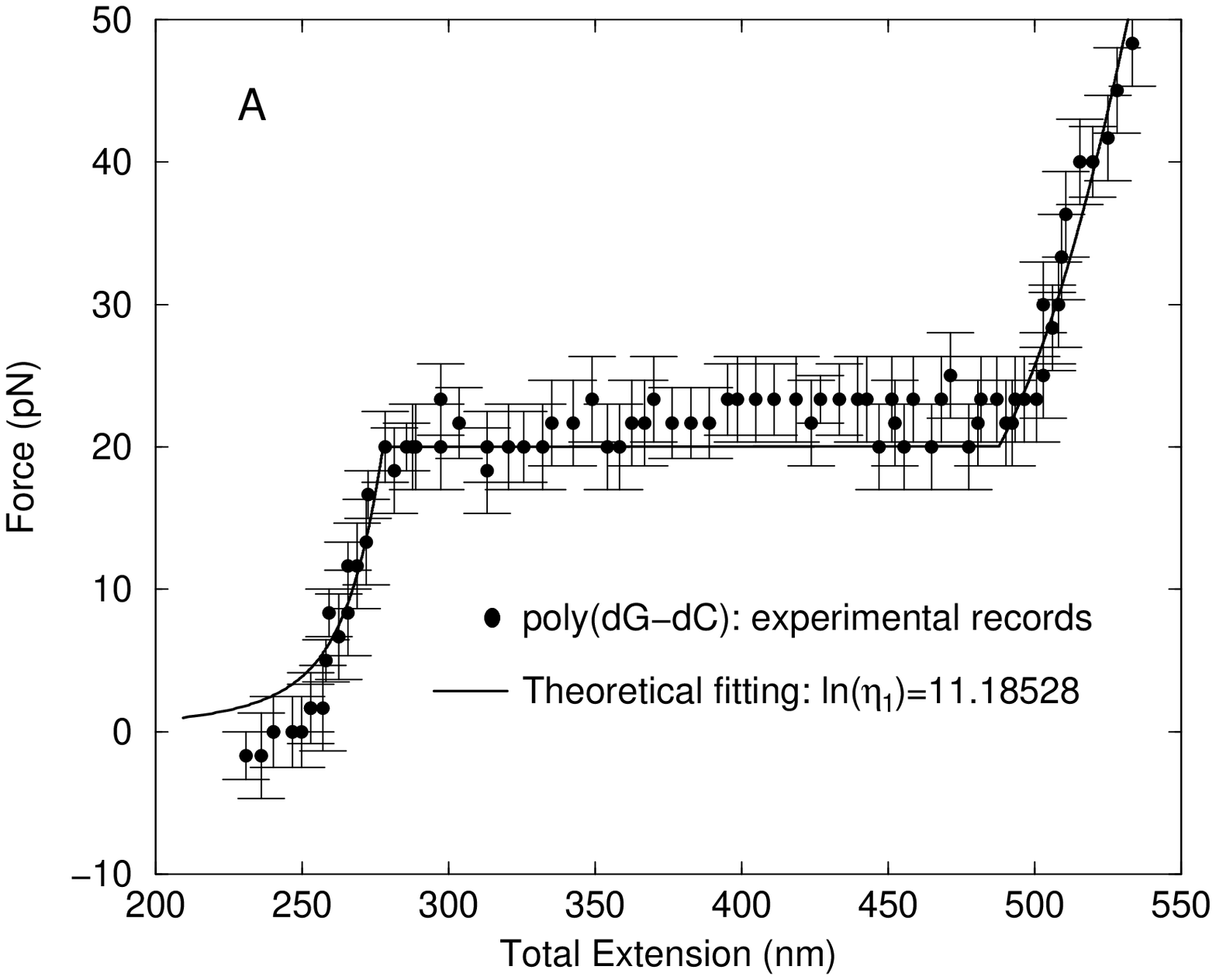,width=8.0cm}

\center 
\psfig{file=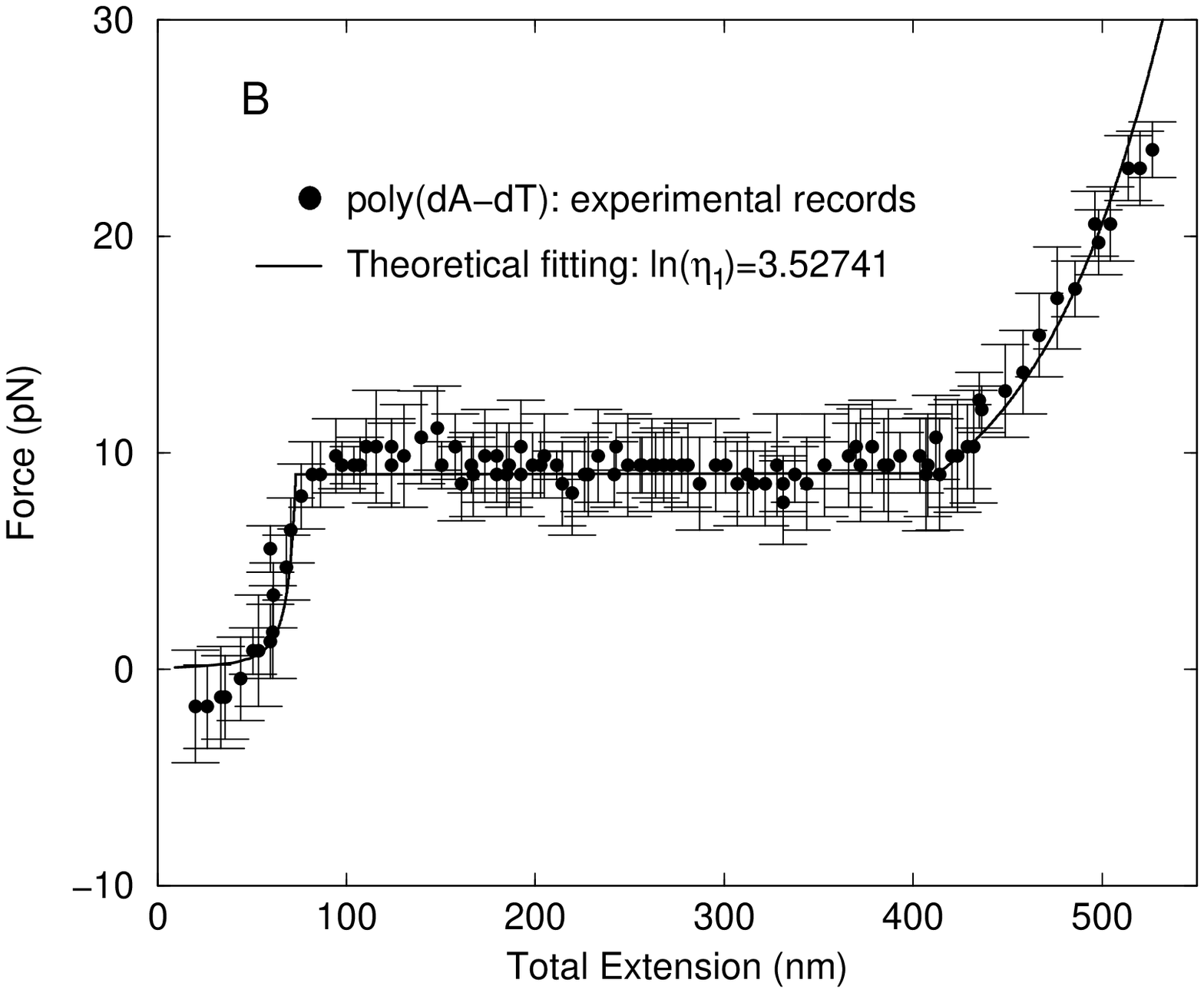,width=8.0cm}
\caption{Force vs extension curves for (A) poly(dG-dC)
and (B) poly(dA-dT) ssDNA chains. The experimental data
are from Ref.~[15] and the theoretical fittings 
are done with {\protect\(b=1.7\protect\)}nm,  
and {\protect\(\ell=1.105 \AA\protect\)} in (A) and 
{\protect\(\ell=1.530 \AA\protect\)}
in (B). The parameter {\protect\(\ln \eta_1\protect\)}
(see Eq.~\ref{eq:eq09}) is determined by the transition force,
and is set  to {\protect\(11.18528\protect\)} and 
{\protect\(3.52741\protect\)},  respectively. To account
for the low extension data a segment of double-stranded
DNA is included in the fitting (since the experiment was
performed by inserting a ssDNA segment in between two
dsDNA segments). 
\label{fig:fig05} }
\end{figure}

In another experiment, Rief {\it et al.}\cite{rief1999} synthesized
single-stranded poly(dG-dC) and poly(dA-dT) DNA chains and investigated
their elastic responses under external force field. They observed 
a highly cooperative elongation process  in both these two
kinds of sequences, with the transition force being $20$ pN for
poly(dG-dC) and $9$ pN for  poly(dA-dT). Figure \ref{fig:fig05}A and 
Fig.~\ref{fig:fig05}B shows the experimental records as well as the
theoretical fittings based on the present model. 

In the case of
poly(dG-dC), the theoretical curve is obtained by calculating the
total extension    of a polymer connected by a dsDNA segment of 
$290.0$nm (with persistence length $53.0$nm and stretch modulus
$S=1000.0$pN as determined by previous experiments\cite{bustamante2000}) 
and a ssDNA segment of
$230.0$nm (with Kuhn length $1.7$nm and bond length variance
$1.105\AA$ as determined by the data of Fig.~\ref{fig:fig04}).
(We have included a segment of dsDNA simply because the experiment
in [15] was performed by inserting a ssDNA segment between two
dsDNA segments).   In the hairpinned state at zero force, 
the average  free energy per Kuhn length  is thus determined by the 
transition force to be $\phi=-(1/2\beta)\ln(\eta_1)=-5.59 k_B T$. 
Similarly, in the case of poly(dA-dT), the fitting is obtained 
with  a  dsDNA segment of $80.0$nm 
and a ssDNA segment of $445.0$nm ($b=1.7$nm, $\ell=1.53$$\AA$)
with the average free energy per Kuhn length being $\phi=-1.763 k_B T$.

Assuming each Kuhn length contains three nucleotide bases (i.e., each
base in the model corresponds to three realistic  nucleotides),
we can infer that the free energy per base-pair in the poly(dG-dC) 
chain is $3.72 k_B T$, while in the poly(dA-dT) chain is
$1.18 k_B T$. These values are close to what we estimated earlier\cite{zhou2001}
and are close to the 
phenomenological parameters $E_{G-C}\sim 3.0 k_B T$ and
$E_{A-T}=1.3 k_B T$ chosen by Bochelmann {\it et al.}\cite{bockelmann1997} to interpret
their experimental results of  separating the
two strands of a dsDNA apart by mechanical force.

\section{Conclusive remarks}
\label{sect5}

In this paper, we have presented an elastic model for single-stranded
DNA and RNA polymers, where both the base-pairing and base-pair stacking
interactions where considered. The theoretical calculations demonstrated that,
depending on the intensity of base-pair stacking potential, ssDNA or
RNA can form two kinds of hairpinned structures, hairpin-I and hairpin-II,
with the base-pairs in hairpin-I only slightly stacked and those in
hairpin-II almost completely stacked. The force-induced hairpin-coil
transition is a second-order process if  the polymer was originally in
hairpin-I macro-state  and a first-order process if it
was in hairpin-II macro-state. The phase diagram and the force-extension
relationship for this polymer system were obtained and the theoretical
results achieved good agreement with experimental observations. 
This work indicates the significance of base-pair stacking interaction
to the structural property of polynucleotide chains.
The existence of such a hairpin-II macro-state in a designed polynucleotide
also justified our previous phenomenological work on the same system
by considering only single-looped hairpin structures.\cite{zhou2001}

The present work has the following implications. 
Given a polynucleotide chain, if the nucleotide bases are randomly
arranged, then the possibility to forming stacked patterns of base-pairs
is very small, since even forming the  smallest stack of two base-pairs 
requires the correlations among four nucleotide bases. Therefore, in
random polynucleotide chains, stacking potential is not important. 
The chains can only form loose hairpins (hairpin-I), 
and structural fluctuations
between these hairpins may be large and quick,   making it difficult
for the existence of any well-defined stable native configurations. 
However, things may be dramatically changed if the polynucleotide
chain composed of the same bases as the random chain rearranges these
nucleotide bases carefully. If the bases in the chain is arranged
in such a way that forming of base-pairs results in stacking of 
base-pairs, then it is very likely that  the polymer will 
fold into certain stable native structures  which are 
much lower in structural  energy than other configurations.
Is this part of the reason why some 
polynucleotide sequences are  folded in a particular manner 
(such as transfer RNA or t-RNA\cite{watson1987})?
The nucleotide sequences in RNA molecules 
of biological cells are the results of millions of years of natural
evolution and selection, and therefore 
in some sense,  they should  all be well designed.

Finally, we just mention that the theoretical method used in
this work to  count for the long-range interactions in ssDNA 
may be applied to investigate the tertiary structures (supercoils)
in double-stranded DNA caused by the topological constraint of
fixed linking numbers. The topological constraint leads to  a 
long-range interaction along the dsDNA chain and  make the
polymer to fold into compacted plectonemic structures.\cite{marko1995b}     
Similar force-extension profiles as the theoretical
curves in Fig.~\ref{fig:fig04} have  been observed.\cite{strick1996}

\section{Acknowledgment}
\label{sect6}

We are grateful to Prof. Z.-C. Ou-Yang. One of the 
authors (H.Z.) appreciates  
a helpful correspondence with A. Montanari,  the  helps of 
U. Bastolla and Jian-Jun Zhou,  as well as the financial 
support of the Alexander von Humboldt Foundation.
H.Z. is also grateful to Prof. R. Lipowsky for a critical
reading of the manuscript and comments.

\end{multicols}

\begin{thebibliography}{99}
\bibitem{note:no1}Since this work is concerned with the 
macroscopical elastic property of polynucleotides and each fictitious
base here represents an average of several real bases of
the polynucleotide, this can be a well justified treatment.
\end{thebibliography}
\end{document}